\documentstyle[12pt]{article}
\def\lsim{\ {}_{\displaystyle \sim}^{\displaystyle <} \ }
\def\Dt{\Delta t}

\begin{document}
\sloppy
\hbox{}
\bigskip
\hspace{6cm} Submitted to {\it Theor. Math. Phys.}
\vspace{2cm}
\begin{center}
{\large ASYMPTOTICS OF EXPANSION OF THE EVOLUTION OPERATOR KERNEL
IN POWERS OF TIME INTERVAL~$\Delta t$ }
\end{center}
\medskip
\centerline{\large V. A. Slobodenyuk }
\vspace{1.5cm}
\begin{center}
{\it Physical--Technical Department, Moscow State University Branch,\\
L. Tolstogo str. 42, 432700 Ulyanovsk, Russia \\
E-mail: slob@ftf.univ.simbirsk.su }
\end{center}
\vspace{3cm}

\centerline{\bf Abstract}

The upper bound for asymptotic behavior of the coefficients
of expansion of the evolution operator kernel in powers of the
time interval $\Dt$ was obtained. It is found that for the
nonpolynomial potentials the coefficients may increase as $n!$.
But increasing may be more slow if the contributions with
opposite signs cancel each other. Particularly, it is not
excluded that for number of the potentials the expansion is
convergent. For the polynomial potentials $\Dt$-expansion
is certainly asymptotic one. The coefficients increase in this
case as $\Gamma(n \frac{L-2}{L+2})$, where
$L$~is the order of the polynom. It means that the point $\Dt=0$
is singular point of the kernel.

\newpage

\hbox{}
\bigskip

\section{Introduction}

Since it was established, that the series of conventional perturbation
theory in coupling constant for anharmonic oscillator and for
number of models in quantum field theory are asymptotic ones
\cite{BW}--\cite{L}, many attempts to construct some other
representations, which would allow to take into account
nonperturbative in coupling constant effects (see, e.~g.,
\cite{HS}--\cite{YY}), were undertaken. Other direction of
research is connected with attempts to obtain physically interesting
information from analysis of divergent series by means of
Borel summation method, Pad\'e approximants etc. (summation of
asymptotic series)~\cite{GGS}--\cite{ZJ}. Presence of a lot of
different approaches means that the problem of nonperturbative
calculations has not satisfactory solution till now.

In paper~\cite{S} attempt was made to calculate the evolution
operator kernel in a form of the series in powers of time interval
$\Dt$. In this paper we examine the asymptotic behavior of that
expansion.

Let us remind briefly the main relations, determining the expansion
of the kernel in powers of $\Dt$~\cite{S}. If the kernel
$\langle q',t'\mid q,t \rangle$ satisfies the Schr\"odinger equation
  \begin{equation} \label{f1}
  i \frac{\partial}{\partial t} \langle q',t'\mid q,t \rangle =
  -\frac{1}{2} \frac{\partial^2}{\partial q'^2}
  \langle q',t'\mid q,t \rangle +
  V(q') \langle q',t'\mid q,t \rangle
  \end{equation}
(here and everywhere below we use dimensionless variables,
defined in standard manner, and for the sake of simplicity consider
only one--dimensional case), then it may be written in a form
  \begin{equation} \label{f2}
  \langle q2',t'\mid q,t \rangle =
  \left( \frac{-i}{2\pi\Delta t} \right)^{1/2}
  \exp \left\{i \frac{(q'-q)^2}{2\Delta t} \right\}
  \left\{1- \sum_{n=1}^{\infty}(i \Delta t)^n Y_n(q',q) \right\}
  \end{equation}
($Y$-representation), or
  \begin{equation} \label{f3}
  \langle q',t'\mid q,t \rangle =
  \left( \frac{-i}{2\pi\Delta t} \right)^{1/2}
  \exp \left\{i \frac{(q'-q)^2}{2\Delta t} -
  \sum_{n=1}^{\infty}(i \Delta t)^n P_n(q',q) \right\}
  \end{equation}
($P$-representation). For the functions $P_n$ and $Y_n$ we have
the recurrent relations
  \begin{equation} \label{f4}
  Y_1(q',q)+(q'-q) \frac{\partial Y_1(q',q)}{\partial q'} =V(q'),
  \end{equation}
  \begin{equation} \label{f5}
  nY_n(q',q)+(q'-q) \frac{\partial Y_n(q',q)}{\partial q'} =
  \frac{1}{2} \frac{\partial ^2 Y_{n-1}(q',q)}
  {\partial q'^2} - V(q')Y_{n-1}(q',q),
  \end{equation}
  \begin{eqnarray} \label{f6}
  &&nP_n(q',q)+(q'-q) \frac{\partial P_n(q',q)}{\partial q'}=
  \\
  &&=\frac{1}{2} \frac{\partial ^2 P_{n-1}(q',q)}{\partial q'^2}-
  \frac{1}{2} \sum_{n'=1}^{n-2}
  \frac{\partial P_{n'}(q',q)}{\partial q'}
  \frac{\partial P_{n-n'-1}(q',q)}{\partial q'}.
  \nonumber
  \end{eqnarray}
And besides $P_1(q',q)=Y_1(q',q)$, $P_1(q',q')=V(q')$. So, we can
calculate the coefficient functions $Y_n$ and $P_n$ till any order
$n$ with a help of~(\ref{f4})--(\ref{f6}) proceeding from given
potential $V(q)$ and then substitute them into~(\ref{f2}),
(\ref{f3}). That will be formal solution of evolution
equation~(\ref{f1}).

The solutions of equations~(\ref{f4})--(\ref{f6}) may be represented
in integral form
  \begin{equation} \label{f7}
  Y_n(q',q)= \int \limits_0^1 \eta^{n-1} d\eta \Phi_n(q+(q'-q)\eta,q),
  \end{equation}
  \begin{equation} \label{f8}
  P_n(q',q)= \int \limits_0^1 \eta^{n-1} d\eta F_n(q+(q'-q)\eta,q),
  \end{equation}
where
  \begin{equation} \label{f9}
  \Phi_n(x,q)= \frac{1}{2} \frac{\partial ^2 Y_{n-1}(x,q)}
  {\partial x^2} - V(x)Y_{n-1}(x,q)
  \end{equation}
and
  \begin{equation} \label{f10}
  F_n(x,q)=
  \frac{1}{2} \frac{\partial ^2 P_{n-1}(x,q)}{\partial x^2}-
  \frac{1}{2} \sum_{n'=1}^{n-2}
  \frac{\partial P_{n'}(x,q)}{\partial x}
  \frac{\partial P_{n-n'-1}(x,q)}{\partial x}
  \end{equation}
are the right hand sides of the equations~(\ref{f5}) and~(\ref{f6}),
respectively. If $n=1$, then $F_1(x,q)=\Phi_1(x,q)=P_1(x,x)=V(x)$
(this is right hand side of~(\ref{f4})).

We  will  use   representations~(\ref{f7})--(\ref{f10})   to
analyze
convergence of the series in~(\ref{f2}), (\ref{f3}).

\section{Study of asymptotics for $Y$-representation}

$Y$-representation is more simple for analysis, because $Y_n$
for given potential $V$ depends on $Y_{n-1}$ only, but not on all
previous coefficient functions $Y_i$ with $i<n-1$, as it take
place for $P$-representation. Let us consider in the beginning
that the potential $V(x)$ is nonpolynomial and all its derivatives
are not equal to zero identically.

We can derive from~(\ref{f7}),~(\ref{f9})
  \begin{eqnarray} \label{f11}
  &&Y_n(q',q)= \int\limits_0^1 d\eta_1 \int\limits_0^1 \eta_2 d\eta_2
  \int\limits_0^1 \eta_3^2 d\eta_3 \dots
  \int\limits_0^1 \eta_n^{n-1} d\eta_n \times
  \\ &&\times
  \left(-V(x_n)+\frac{1}{2} \frac{\partial^2}{\partial x_n^2}\right)
  \left(-V(x_{n-1})+\frac{1}{2}
                 \frac{\partial ^2}{\partial x_{n-1}^2}\right) \dots
  \left(-V(x_3)+\frac{1}{2} \frac{\partial^2}{\partial x_3^2}\right)
  \times \nonumber \\ &&\times
  \left(-V(x_2)+\frac{1}{2} \frac{\partial^2}{\partial x_2^2}\right)
  V(x_1).
  \nonumber
  \end{eqnarray}
Here for brevity we denoted $x_i=q+(x_{i+1}-q)\eta_i$.
Generally, $x_i$ with different numbers $i$ are connected as
follows
  \begin{equation} \label{f12}
  x_i=q+(x_{i+l}-q)\eta_i \eta_{i+1} \dots \eta_{i+l-1},
  \end{equation}
and derivatives, as it is seen from~(\ref{f12}), are connected by
the relations
  \begin{equation} \label{f13}
  \frac{\partial}{\partial x_i}=\eta_{i-l} \eta_{i-l+1} \dots
  \eta_{i-1} \frac{\partial}{\partial x_{i-l}}.
  \end{equation}

Disclosing brackets in~(\ref{f11}) and reducing with a help
of~(\ref{f13}) all differentiations of the potential to
differentiation of it with respect to full argument, i.~e. to the
form $V^{(k)}(x_i) \equiv \frac{d^k}{d x_i^k}V(x_i)$, we can
get the expression for $Y_n$ with following structure
  \begin{eqnarray} \label{f14}
  Y_n(q',q)&=& \int\limits_0^1 d\eta_1 \int\limits_0^1 \eta_2 d\eta_2
  \int\limits_0^1 \eta_3^2 d\eta_3 \dots
  \int\limits_0^1 \eta_n^{n-1} d\eta_n \times
  \\ &&\times
  \left\{ \sum_{k=1}^n \sum_{\{l_i\}} \kappa_{l_1,\dots,l_k}^{(k)}
  V^{(l_1)} \dots V^{(l_k)} \right\}.
  \nonumber
  \end{eqnarray}
The sum $\sum \limits_{\{l_i\}}$ is taken over that sets of $l_i$,
for which total order of derivatives is
$\sum \limits_{i=1}^k l_i =2(n-k)$. Besides, one is to take into
account the terms with different arguments of function $V^{(l_i)}$.
The coefficients $\kappa ^{(k)}$ contain products of $\eta_i$.

Let us estimate $Y_n$ using~(\ref{f14}) and taking into account
that, according to~(\ref{f11}), the argument of the potential $V$
belongs to the interval $[q,~q']$. If we do not consider the singular
potentials, then we can adopt that at finite interval every function
$V(x)$ with all its derivatives is bounded, i.~e. for every number
$k$ the positive constant $C_k$ exists, so that for all $x$ from the
interval $[q,~q']$ following condition is satisfied:
 $|V^{(k)}(x)| < C_k$. Then we have
  \begin{eqnarray} \label{f15}
  &&|Y_n(q',q)| \leq
  \\
  &&\leq \int\limits_0^1 d\eta_1 \int\limits_0^1 \eta_2 d\eta_2
  \dots \int\limits_0^1 \eta_n^{n-1} d\eta_n
  \left\{ \sum_{k=1}^n \left|
  \sum_{\{l_i\}} \kappa_{l_1,\dots,l_k}^{(k)}
  V^{(l_1)} \dots V^{(l_k)} \right| \right\} <
  \nonumber \\
  &&< \int\limits_0^1 d\eta_1 \int\limits_0^1 \eta_2 d\eta_2
  \dots \int\limits_0^1 \eta_n^{n-1} d\eta_n
  \left\{ \sum_{k=1}^n C^k \left|
  \sum_{\{l_i\}} \kappa_{l_1,\dots,l_k}^{(k)} \right| \right\}.
  \nonumber
  \end{eqnarray}
In~(\ref{f15}) we denoted $C \equiv \max \limits_{\{k\}}C_k$.
It is hard to calculate
$\sum \limits_{\{l_i\}} \kappa_{l_1,\dots,l_k}^{(k)}$ exactly,
but it is possible to estimate it from above, putting in
$\kappa^{(k)}$ all $\eta_i$ to be equal to $1$. Then the
coefficients $N(n,k)= \sum \limits_{\{l_i\}}
 \kappa_{l_1,\dots,l_k}^{(k)}(\eta_i=1)$
can be calculated recurrently from the equation
  \begin{equation} \label{f16}
  N(n,k)= N(n,k-1)+\frac{1}{2}k^2 N(n-1,k),
  \end{equation}
which is direct consequence of~(\ref{f7}), (\ref{f9}).
Here $N(n,k)$ is different from zero only if $k$ varies from $1$
to $n$, besides $N(1,1)=1$.

It is easy to see, that running recurrent equation~(\ref{f16})
from up to down we get  the main contribution for large $n$ from
the terms, corresponding to folloing path of transition from $N(n,k)$
to $N(1,1)$: at first we should fix $k$ and decrease $n$ by one
unit per step till $n$ becomes equal to $k$, then we should
decrease both $k$ and $n$ by one unit per step till $n$ and $k$
become equal to $1$. This leads to estimate
  \begin{equation} \label{f17}
  N(n,k) \sim \left( \frac{k^2}{2} \right)^{n-k}.
  \end{equation}
Let us find the number $k$, corresponding to the terms, giving
the main contribution to~(\ref{f14}) at large $n$. For this we
test the right hand side of~(\ref{f17}) for an extremum as function
of  $k$  (temporary  we  consider   $k$   real)   at   fixed
sufficiently
large $n$. The value $k_m=[k]$ ($[\dots]$ is integer part of number),
where $k$ is the root of equation
  \begin{equation} \label{f18}
  k(2\log k + 2-\log 2)=2n,
  \end{equation}
corresponds to maximum contribution. Equation~(\ref{f18}) has one
root, which may be calculated approximately with a help of
relation
  \begin{equation} \label{f19}
  k \approx \frac{n}{\log n}\left(1+\frac{\log \log n}{\log n} \right).
  \end{equation}
So, the main contribution is determined by the coefficient
  \begin{equation} \label{f20}
  N(n,k_m) \sim \frac{n^{2n}}{(2e^2)^n(\log n)^{2n}}.
  \end{equation}
We can now estimate $Y_n$ proceeding from~(\ref{f20}) and taking into
consideration that integration over $\eta_i$ in~(\ref{f15}) gives the
factor $1/n!$:
  \begin{equation} \label{f21}
  |Y_n(q',q)| \lsim \frac{1}{n!} C^{k_m} N(n,k_m) \sim
  \frac{n!}{n} \frac{C^{\frac{n}{\log n}}}{2^n (\log n)^{2n}}
  \sim \frac{n!}{n 2^n(\log n)^{2n}}.
  \end{equation}
This evaluation is overestimated, because we put $\eta_i=1$ in
$\kappa ^{(k)}$ and because we do not consider in~(\ref{f14})
possible cancellations of the terms, corresponding to the same
number of multipliers $k$. Cancellations may occur thanks to
different signs of derivatives $V^{(l_i)}(x)$ for different orders
$l_i$. Uncertainty of the estimate of the first kind can be
evaluated from consideration in~(\ref{f14}) the terms of form
  \begin{displaymath}
  \frac{\partial^{2(n-1)}V(x_1)}{\partial x_n^2
  \partial x_{n-1}^2 \dots \partial x_2^2}=
  \eta_{n-1}^2 \eta_{n-2}^4 \cdots \eta_1^{2n} V^{(2(n-1))}(x_1),
  \end{displaymath}
which give the minimal contribution when dependence of
$\kappa ^{(k)}$ on $\eta_i$ is taken into account. Instead of
factor $1/n!$ in~(\ref{f21}) that term gives
  \begin{displaymath}
  \frac{n!}{(2n)!} \sim \frac{\sqrt{n}}{n! 4^n}.
  \end{displaymath}
So, correct estimate lies at the bounds
  \begin{equation} \label{f22}
  |Y_n(q',q)| \lsim \left( \frac{\sqrt{n}}{4^n} \div 1 \right)
  \frac{n!}{n} \frac{C^{\frac{n}{\log n}}}{2^n (\log n)^{2n}}.
  \end{equation}
This  correction does not touch the main factor
 ${\displaystyle \frac{n!}{(\log n)^{2n}}}$,
which means, that if contributions of different signs does not
cancel each other essentially, then the representation~(\ref{f2})
is asymptotic one. Nevertheless, we do not state, that for every
potential expansion will be divergent. On the contrary, one may
hope, that that functions $V(x)$ exist (besides squared potentials),
for which in terms in~(\ref{f14}) essential cancellations take
place and the coefficient functions $Y_n$ rise more slowly,
then $n!$.

\vspace{1.cm}

Now we consider the polynomial potentials. It was supposed in the
estimates made above, that derivatives of the potentials of all
orders do not equal to zero identically. This is not so for
polynomial potentials. So, it is  necessary  to  modify  our
reasonings.

Let $V(x)$ is polynom of order $L$. Then $V^{(L+1)}(x) \equiv0$.
We will consider the terms in~(\ref{f14}) containing $k$
multipliers $V^{(l_i)}$. Because $\sum \limits_{i=1}^kl_i=2(n-k)$
and the order of derivative, acting on every multiplier, is not
higher then $L$, then this term is not equal to zero only if
$2(n-k) \leq kL$ or $k \geq {\displaystyle \frac{2n}{L+2}}$.
So, there exists the down bound for possible values of $k$, which
is higher, at large $n$, then $k_m$ obtained from~(\ref{f19}). In
case of the polynomial potential one has to take $k_m$ in a form
  \begin{equation} \label{f23}
  k_m = \left[ \frac{2n}{L+2} \right] .
  \end{equation}
Then we find from~(\ref{f17}) taking into account~(\ref{f23})
  \begin{equation} \label{f24}
  N(n,k_m) \sim \sqrt{n} \left( \frac{e^2}{2L^2} \right)^
  {\frac{nL}{L+2}} \Gamma \left( n \frac{2L}{L+2} \right).
  \end{equation}
And for $Y_n$ we have estimate, founded on~(\ref{f24}):
  \begin{equation} \label{f25}
  |Y_n(q',q)| \lsim \left(
  \frac{(2e^2)^LC^2}{(L-2)^{(L-2)}(L+2)^{(L+2)}}
  \right)^{\frac{n}{L+2}} \Gamma \left( n \frac{L-2}{L+2} \right).
  \end{equation}

It turned out to be that for the polynomial potentials the
representation~(\ref{f2}) is asymptotic one. But the coefficient
functions rise with growth of $n$ more slowly, then $n!$.
Particularly, for $L=4$ we have from~(\ref{f25})
$|Y_n| \sim \Gamma(n/3)$. For harmonic oscillator, as it is naturally,
$|Y_n| \sim \left( {\displaystyle \frac{e}{4}} \sqrt{2C} \right)^n$,
i.~e. the series in~(\ref{f2}) has finite convergence range.

\section{Asymptotics for $P$-representation }

Let   us   consider   now   the   representation~(\ref{f3}).
Analogously
to previous case we introduce the coefficients $N(n,k)$, which
satisfy the relation
  \begin{eqnarray} \label{f26}
  &&N(n,k)= \frac{1}{2}k^2 N(n-1,k) +  \\
  &&+\frac{1}{2} \sum_{n'=1}^{n-2} \sum_{k'=1}^{k-1} k'(k-k')
  N(n',k') N(n-n'-1,k-k'),        \nonumber
  \end{eqnarray}
following from~(\ref{f18}). Here $N(1,1)=1$ and $N(n,k)$ is
different from zero only if $1 \leq k \leq \left[
\frac{n+1}{2} \right]$. The main contribution at large $n$ comes
from the terms, which arise in running of~(\ref{f26}) from up to
down when the numbers $n,~k$ are changed in following manner:
at the beginning we fix $k$ and decrease $n$ till $n$ becomes
equal to $2k$, then we decrease $n$ by two units and $k$ by one unit
per step till $n$ and $k$ become equal to $1$. It gives:
  \begin{equation} \label{f27}
  N(n,k) \sim \frac{k^{3/2}}{2^{n-k}} k^{2n-3k}.
  \end{equation}
Maximal contribution will be got from the terms, containing
$k_m=[k]$ multipliers $V^{(l_i)}$, where $k$ is determined by
equation
  \begin{equation} \label{f28}
  k(3\log k + 3-\log 2)=2n.
  \end{equation}
Equation~(\ref{f28}) has only one root, which is approximately
(for large $n$) equal to
  \begin{equation} \label{f29}
  k \approx \frac{2}{3} \frac{n}{\log n}
  \left(1+\frac{\log \log n}{\log n} \right).
  \end{equation}
Then asymptotically main contribution is determined, as is seen
from~(\ref{f27}), (\ref{f29}), by the coefficient
  \begin{equation} \label{f30}
  N(n,k_m) \sim \frac{n^{2n+3/2}}{(2e^2)^n (\log n)^{2n}}.
  \end{equation}
When we considered $Y$-representation, calculation of the
integrals of kind $\int \limits_0^1 \eta_i^{i-1} d\eta_i$ led to the
factor $1/n!$ (see ~(\ref{f15}), (\ref{f21})).
In this case situation is rather complicated. When we express by
means of~(\ref{f6}) the coefficient $P_n(q',q)$ through lower
coefficient functions, decreasing $n$ as it was described above,
then the integral
  \begin{eqnarray*}
  &&\int\limits_0^1 \eta_n^{n-1} d\eta_n
  \int\limits_0^1 \eta_{n-1}^{n-2} d\eta_{n-1} \dots
  \int\limits_0^1 \eta_{2k_m}^{2k_m-1} d\eta_{2k_m} \times
  \\ &&\times
  \int\limits_0^1 \eta_{2k_m-1}^{2k_m-2} d\eta_{2k_m-1}
  \int\limits_0^1 \eta_{2k_m-3}^{2k_m-4} d\eta_{2k_m-3} \dots
  \int\limits_0^1 \eta_3^2 d\eta_3
  \int\limits_0^1 \eta_1^0 d\eta_1
  \end{eqnarray*}
arises. It gives for estimate of $P_n$ the factor
  \begin{equation} \label{f31}
  \frac{2^{k_m-1} (k_m-1)!}{n!}.
  \end{equation}
Involving~(\ref{f31}) we get following evaluation
  \begin{equation} \label{f32}
  |P_n(q',q)| \lsim \frac{2^{k_m}k_m!}{n!} C^{k_m} N(n,k_m) \sim
  \frac{n^{1/2}n!}{(\log n)^{2n}} \left( \frac{e^{2/3}}{2} \right)^n.
  \end{equation}
This evaluation, so as~(\ref{f21}), is overestimated. Asymptotic
increase of type $n!$ will take place only if there is no
essential cancellations of the contributions with different signs.

\vspace{1.cm}

Consider the potentials which are the polynoms of order $L$.
In this case almost all previous reasonings are right, excluding
one point. For $k_m$ one should take expression~(\ref{f23})
instead of~(\ref{f29}). Then we have evaluation
  \begin{equation} \label{f33}
  |P_n(q',q)| \lsim n^2 2^n \left( \frac{e^{2(L-1)}C^2}
  {(L-2)^{(L-2)}(L+2)^{(L+2)}} \right)^{\frac{n}{L+2}}
  \Gamma \left( n \frac{L-2}{L+2} \right).
  \end{equation}
The main asymptotic behavior in~(\ref{f33}), so as in~(\ref{f25}),
is determined by the factor
$\Gamma \left( n \frac{L-2}{L+2} \right)$. So, for $L>2$ the
series in $P$-representation are asymptotic one too.

Note, that in estimates~(\ref{f25}),~(\ref{f33}) for the polynomial
potentials, maybe, the factors of type $n^a b^n$, are some
overestimated, but the main rise of type
$\Gamma \left( n \frac{L-2}{L+2} \right)$ is established exactly.
For the nonpolynomial potentials estimates (\ref{f21}), (\ref{f22}),
(\ref{f32}) determine only upper bounds for possible asymptotic
rise. They can be not achieved for some potentials.

\section{Conclusion}

As it is seen from~(\ref{f25}),~(\ref{f33}), in the case of
polynomial potentials the coefficient functions in expansions
(\ref{f2}),~(\ref{f3}) rise as
$n^ab^n \Gamma \left( n \frac{L-2}{L+2} \right)$ for
$n \to \infty$. So, for $L>2$ the series in ~(\ref{f2}),
(\ref{f3}) are asymptotic  ones.  We  can  note  here,  that
divergence
of expansion in $\Dt$ means that the point $\Dt =0$ is singular
point of the kernel. Discussion of this topic can be found in
\cite{S3}.

If the potential is nonpolynomial continuous function, then for
asymptotics of the coefficient functions one has (see~(\ref{f22}),
(\ref{f32})) an upper bound of kind
$n^a b^n {\displaystyle \frac{n!}{(\log n)^{2n}}}$. Nevertheless,
this bound was obtained by summation of all contributions
without taking into account the signs of the terms. So, real
asymptotic rise for some potentials may be essentially more
slow. Particularly, possibility of existing of number potentials,
for which the expansion in $\Dt$ is convergent, is not excluded.

Note, that all presented reasonings may be easily generalized on
many-dimensional case, because the representations of
type (\ref{f2})--(\ref{f10}) for $D$-dimensional space can be
derived from ones for one-dimensional space by changing of
one-dimensional space variables to vector variables and by
changing of derivatives over coordinates to $D$-dimensional
operator {\it grad}. It leads to obvious changes in equations
(\ref{f16}),~(\ref{f26}) for estimates of asymptotics:
instead of factor $k^2$ one should take $Dk^2$, and instead
of $k'(k-k')$ one should take $D^2 k'(k-k')$. That substitution
modifies final results not essentially: in~(\ref{f20}), (\ref{f21}),
(\ref{f22}), (\ref{f24}), (\ref{f25}), (\ref{f30}), (\ref{f32}),
(\ref{f33}) one is to add the factor $D^n$. It does not affect
the qualitative results obtained above.

In conclusion, let us pay attention to one feature of the
representations~(\ref{f2})--(\ref{f10}). When we calculate the
transition matrix element
$\langle \vec q\,',t'\mid \vec q,t \rangle$, only values of the
potential $V(\vec x)$ in the points, lying on the segment,
joining the points $\vec q$ and $\vec q\,'$, are essential.
Behavior of the potential in other points does not affect on
value of the kernel at all!


\vspace{2cm}

\begin{thebibliography} {99}

\bibitem{BW}
{\it Bender C. M., Wu T. T.} // Phys.~Rev.~Lett. 1971. N.~7.
V.~27. P.~461--465.

\bibitem{BW1}
{\it Bender C. M., Wu T. T.} // Phys.~Rev. 1971. V.~D7. N.~6.
P.~1620--1636.

\bibitem{L}
{\it Lipatov L. N. } // JETF. 1977. V.~72. N.~2. P.411--427.

\bibitem{HS}
{\it Halliday I. G., Suranyi P.} // Phys.~Rev. 1980. V.~D21.
N.~6. P.~1529--1537.

\bibitem{U}
{\it Ushveridze A. G. } // Yad. Fiz. 1983. V.~38. N.~3(9).
P.~798--809.

\bibitem{CC}
{\it Consoli M., Ciancitto A.} // Nucl.~Phys. 1985. V.~B254.
N.~3,4. P.~653--677.

\bibitem{NSR}
{\it Namgung W., Stevensin P. M., Reed J. F.} //
Z.~Phys.~C~---~Particles and Fields. 1989. V.~45. N.~1. P.~47--56.

\bibitem{BMPS}
{\it Bender C. M., Milton K. A., Pinsky S.S., Simmons L. M., Jr.} //
J.~Math.~Phys. 1990. V.~31. N.~11. P.~2722--2725.

\bibitem{BBL}
{\it Bender C. M., Boettcher S., Lipatov L. N.} // Phys.~Rev.~Lett.
1992. V.~68. N.~25. P.~3674--3677.

\bibitem{SS}
{\it Sissakian A. N., Solovtsov I. L.} //
Z.~Phys.~C~---~Particles~and~Fields. 1992. V.~54. N.~2. P.~263--271.

\bibitem{SSS}
{\it Sissakian A. N., Solovtsov I. L., Shevchenko O. Yu.} //
Phys.~Lett. 1992. V.~297B. N.~3,4. P.~305--308.

\bibitem{YY}
{\it Yukalova E. P., Yukalov V. I.} // Phys.~Lett. 1993. V.~175A.
2N.~1. P.~27--35.

\bibitem{GGS}
{\it Graffi S., Grecchi V., Simon B.} // Phys.~Lett. 1970. V.~32B.
N.~7. P.~631--634.

\bibitem{KTS}
{\it Kazakov D. I., Tarasov O. V., and Shirkov D. V.} // Yad. Fiz.
1979. V.~38. N.~1. P.~15--25.

\bibitem{KS}
{\it Kasakov D. I., Shirkov D. V.} // Fortschr.~Phys. 1980. V.~28.
P.~465--499.

\bibitem{ZJ}
{\it Zinn-Justin J.} // Phys.~Rep. 1981. V.~70. N.~2. P.~109--167.

\bibitem{S}
{\it Slobodenyuk V. A.} // Z.~Phys.~C~---~Particles~and~Fields.
1993. V.~58. N.~4. P.~575--580.

\bibitem{S3}
{\it Slobodenyuk V. A.} // {\it Singularity of the Evolution
Operator Kernel in Time Variable. } Submitted to Mod. Phys. Lett. A.

\end{thebibliography}
\end{document}